Article

# Implementation of Tools for Lessening the Influence of Artifacts in EEG Signal Analysis

Mario Molina-Molina, Lorenzo J. Tardón *, Ana M. Barbancho and Isabel Barbancho *

ATIC Research Group, ETSI Telecomunicación, Universidad de Málaga, 29071 Málaga, Spain;
mmolina@ic.uma.es (M.M.-M.); abp@uma.es (A.M.B.)
* Correspondence: ltg@uma.es (L.J.T.); ibp@uma.es (I.B.)

**Abstract:** This manuscript describes an implementation of scripts of code aimed at reducing the influence of artifacts, specifically focused on ocular artifacts, in the measurement and processing of electroencephalogram (EEG) signals. This process is of importance because it benefits the analysis and study of long trial samples when the appearance of ocular artifacts cannot be avoided by simply discarding trials. The implementations provided to the reader illustrate, with slight modifications, previously proposed methods aimed at the partial or complete elimination of EEG channels or components obtained after independent component analysis (ICA) of EEG signals. These channels or components are those that resemble the electro-oculogram (EOG) signals in which artifacts are detected. In addition to the description of each of the provided functions, examples of utilization and illustrative figures will be included to show the expected results and processing pipeline.

## 1. Introduction

In most electroencephalogram (EEG) signal recordings, artifacts related to eye movement, eye blinks, muscle movements, heartbeat, or other noises can be found. These produce voltage levels that, together with the desired signals, are picked up by the electrodes, thus contaminating the brain signals. In many analysis contexts, it is desirable to identify these artifacts in order to either eliminate, reduce, or correct them.

Methods for the correction of eye blinks and artifacts such as those based on time or frequency domain regression [1,2] or Kalman filtering [3] can be found in the literature. However, artifacts are mingled with EEG signals in both dimensions: time and frequency. This causes the decontamination or cleaning of EEG signals with regression-based methods to produce loss of relevant information. Some more recent methods are based on mixing algorithms such as principal component analysis (PCA) or independent component analysis (ICA), which are able to separate signals in multichannel EEG recordings [4].

Jiang et al. provide, in [5], a review of artifact removal techniques, which includes regression methods, the use of wavelet transform (WT), and blind source separation schemes, including PCA, ICA, canonical correlation analysis (CCA), empirical mode decomposition schemes, filtering methods, and other hybrid techniques. Among them, ICA, which is our selected analysis tool, seems to have attracted the most attention in the past years, even in other related contexts, such as electrocardiogram signal analyses [6]. In spite of the many existent proposals, the authors still observe artifact removal as an open problem.

Similarly, Vandana Roy et al., in [7], review a wide number of research papers of the same topic, finding a similar set of types of methods to address the artifact removal process, where the electro-oculogram (EOG) artifact is considered the most dominant one.

Gorjan et al. [8] categorize artifact removal methods into those based on filtering, those based on blind source separation, and those based on artifact subspace reconstruction (ASR), which performs PCA decomposition on EEG data windows for artifact component identification and removal. Combined schemes are also observed, which often consider ICA plus signal transforms, such as WT [9].

The utilization of WT is another topic that has attracted some attention for the treatment of artifacts in EEG signals, especially in the specific context of single channel EEG signals [10–12].

From a different perspective, Mumtaz, Rasheed, and Irfan [13] analyze the availability and usability of artifact removal methods from EEG signals, identifying diverse issues that slow the progress of artifact removal techniques; among them, the following were found: lack of open-source EEG artifact datasets, and source implementations.

Looking at the literature reviewed, it is clear that the use of ICA, either solely or in combination with other techniques or modifications, constitutes a solid and interesting base for the development of methods to deal with artifacts in EEG signals. In addition, ICA is seen as one of the reference methods in several stages of our processing context, which makes it of great interest to provide open source code, data, and examples that allow simple interaction with ICA methods, and it also benefits the development of further artifact removal techniques. These are all aspects addressed by this manuscript.

Specifically, in our case, the proposed implementations for artifact pre-processing are linked to the methods that appeared in [14], which perform complete or partial removal of ICA components or excerpts similar to the simultaneous EOG signals responsible for certain artifacts. Nevertheless, the ideas and implementations are extensible to other detectable artifacts.

The elimination of independent components is performed, either completely or partially, with the help of a membership function (MSF) employed to identify artifact events: a vector that identifies samples where an ocular artifact is detected. This MSF vector can be built with the functions proposed here, or by using other artifact detection algorithms and the adaption of their output to the simple format employed in the sample implementation provided.

In the next section, a brief description of the data employed for the illustration of the utilization of the proposed implementation is given; later, a very short reminder of the methods described in [14] is shown. Then, in Section 4, specific details about our implementation are displayed, together with the required parameters and related functions. Some illustrative examples and results are shown in Section 5; finally, the conclusions derived from the work presented in this manuscript are presented.

## 2. EEG Database

In this section, the data employed to illustrate the development, recordings, methodology, equipment, and processing stages are described.

### 2.1. Equipment

The EEG data used in the tests to illustrate the performance and sample results of the algorithms have been obtained by the ATIC research group at Escuela Técnica Superior de Ingeniería de Telecommunication building of Universidad de Málaga. They were acquired using an EasyCap electrode cap (Standard 96Ch actiCAP snap) with 64 channels.

Two of the EEG channels were relocated to capture eye movement in the vertical (VEOG) and horizontal (HEOG) planes. In addition, one of the channels (FPz) acts as the ground, placed in the center of the forehead, and another channel is employed as reference (FCz), located on the mid-line vertex of the head.

The sampling rate is set to 2500 Hz, with the impedance of the electrodes during recording sessions below 10 K$\Omega$; specifically, the impedance during the recording session corresponding to the examples and data shown in this manuscript is between 1 and 8 K$\Omega$.

### 2.2. Experiment

The experiment consisted of recording sessions in which a subject performs four different but related activities: observe and think of certain words (or concepts) in Spanish, see associated symbols, hear words in Spanish, and read aloud selected words in Spanish. A few concepts are employed that are presented by all the stimuli types and activity; e.g., the word 'no' is shown, heard, though of, and read, and a red cross, representing negation, is also shown.

The stimuli consisting of showing words in Spanish, or associated symbols or sounds, last between 1 and 3 s, depending on the different representations of the concept, or their duration; an additional second separates the succeeding activity. The activities of reading and thinking of a certain word last a total of 6 s.

The stimuli or activities, and words, images, and concepts, appear in random order during the session, which lasts around 35 min.

The recording sessions were carried out in a separate room where the test subject was sitting alone on a chair in front of a computer screen that presents the stimuli and activities. The software tool E-prime 3 [15] was employed to present the stimuli and activities to the test subjects.

Prior to carrying out the experiment, the subjects were informed about the different tasks the experiment consisted of, and they were instructed to relax and avoid unnecessary movement during the recording of EEG signals.

Specifically, the results drawn in Section 5 correspond to the stimulus in which the subject visualizes an image related to the meaning of the word 'truck.'

Data acquisition methodology was conducted according to the guidelines of the Declaration of Helsinki and approved by the Comité de Ética de la Investigación Provincial de Málaga on 19 December 2019, approval number: 2176-N-19. Informed consent was obtained from all subjects involved in the study.

### 2.3. EEG Processing

Data were processed using Fieldtrip [16,17], version of 11 December 2022, and Matlab R2022b [18].

Data are pre-processed to correct the trend and the reference level is set to 0. Then, a low-pass filter with cut-off frequency 250 Hz is applied to remove frequencies well over the EEG range; also, a high-pass filter with cut-off frequency 1 Hz is applied to further remove the very-low frequency components. Note that, in our case, the common band-stop filter between 49 and 51 Hz, to remove the power line frequency at 50 Hz, is not necessary since the equipment used for the recordings is battery operated.

The processing scheme is included for illustrative purposes; it should be adapted to the specific data and analysis or research objectives.

## 3. Tools and Methods

As explained previously, the implemented methods are based on [14], although slight modifications have been included in the implementation provided at https://hdl.handle.net/10630/26154 (accessed on 16 January 2024), aimed at encouraging other researchers to follow the techniques described and apply their specific processing decisions. Next, a brief summary of each of them is presented, but first, the idea of artifact membership function (MSF) is outlined.

**Artifact membership function.** The artifact membership function (MSF), inspired by [19], is a logical sequence with the same length of the recorded EEG data, in which artifacts are marked with $+1$ (or a positive value between 0 and 1) and the rest of the samples with 0.

In [14], this function is manually created for the evaluation of the methods presented; however, in this work, to facilitate the utilization of the data and code provided, a function (*fta_create_msf_fieldtrip*) that automatically creates an MSF using Fieldtrip toolbox functions will be described for illustrative purposes. Note that other artifact detection algorithms, such as the one proposed in [20], based on the calculation of the acceleration of EOG signals, or the one described in [21], based on the calculation of probabilistic models from the derivatives of VEOG and HEOG signals, or the procedure described by Rashida and Habib [22], based on the extraction of features from EEG channels regenerated from signal components extracted by ICA, could be equally employed, along with others.

**Method 1: ICs removal.** In this method, the correlations between the components obtained by ICA and EOG signals are obtained with the objective of eliminating the independent components (ICs) that have higher similarity to EOG signals (we consider vertical, VEOG, and horizontal, HEOG, signals). Subsequently, EEG channels are reconstructed without those components. This method can be seen as related to the works in [23], where the author experiments with EEG signal filtering and the removal of independent components (ICs), or [24], where the authors discard ICs on the basis of a Bayesian deep-learning scheme.

**Method 2: Partial removal of ICs.** This method is similar to the previous one, but now the MSF is used. The procedure is as follows: the correlation between the components obtained after ICA with all channels and EOG channels is calculated to identify components similar to the EOG signals. Then, instead of eliminating them, the components are set to zero only in the samples where artifacts appear in order to later reconstruct EEG channels without these artifactual samples.

**Method 3: Partial removal of ICs using an artifact-free ICA unmixing matrix.** In this case, the unmixing matrix for ICA is calculated from EEG signals without the samples with artifacts. Samples with artifacts must be identified by using an MSF. As in the previous methods, the correlation with EOG signals is obtained to select the components with larger values. In this case, as in the previous method, the samples related to a positive value of the MSF signal, i.e., those in which artifacts were detected, will be eliminated.

## 4. Implementation

In this section, each of the functions implemented to apply the methods exposed to EEG data are explained in depth. This implementation is aimed at serving as a guide for the usage of the methods described in [14] and fostering the improvement and implementation of the processing schemes in other software pieces.

### 4.1. Complete Removal of Independent Components: fta_ica_removed_components_method

This function completely removes the ICs of EEG signals with a high correlation with EOG signals. To this end, the correlation coefficients between the ICs of EEG signals and EOG data are obtained, sorted, and the ones with the largest amplitude are removed to obtain the clean data.

The usage of this function is shown next:

*cleaned_data = fta_ICA_removed_components_method(data_eeg, fig)*: this function removes ICs from the input EEG data based on correlation and draws (if *fig* = 1) a figure with the correlation values between the EEG and EOG signals, before and after the process.

This function makes use of the following:

- *ft_selectdata*: Fieltrip function to select EOG data.
- *ft_componentanalysis*: Fieldtrip function that obtains ICs from EEG data for all channels.
- *fta_corrcoeff*: Ad hoc function that calculates correlation coefficients (see Section 4.2).
- *fta_select_comp_from_coeff_correlation*: Ad hoc function to select ICs with a high correlation between EOG signals and ICs (see Section 4.3).
- *fta_channels_from_ica_reject_components*: Ad hoc function to reject components previously selected, e.g., using *fta_select_comp_from_coeff_correlation* (see Section 4.4).

The steps followed by this method are shown next:

1. Obtain the vertical and horizontal electro-oculogram channels: VEOG and HEOG.
2. Obtain the ICs from all EEG channels using the Fieldtrip function *ft_componentanalysis*.
3. Calculate the correlation coefficients between the ICs and EOG signals using the function *fta_corrcoeff* (see Section 4.2). A matrix of coefficients is returned, where each row corresponds to each EOG channel and each column to the ICs of the EEG signals.
4. The sum of all the modulus of correlation coefficients of the EOG channels with each of the independent components is obtained.
5. ICs to be removed are selected using the function *fta_select_comp_from_coeff_correlation* (see Section 4.3). To do this, outliers are selected. In addition, the sum of the correlation coefficients is ordered from highest to lowest to show these components can be selected using a different criterion. These components are the ones selected for elimination.

6. The selected ICs are removed using the function *fta_channels_from_ica_reject_components* (see Section 4.4). This function reconstructs the EEG channels from ICA components, except those selected for removal.

The inputs to this function are as follows:

1. *data_eeg*: Fieldtrip-structured data with EEG data to remove ocular artifacts.
2. *fig*: Integer (optional) scalar value (1 or 0) to plot, for illustration, the sum of the correlation coefficients before and after applying the method.

The output of the function are as follows:

1. *cleaned_data*: Fieldtrip-structured data with less influence from ocular artifacts.

In Section 5, the scripts of the Matlab code will be shown.

*4.2. Similarity between EEG and EOG Data: fta_corrcoeff*

This function calculates the correlation coefficients between EOG and EEG channels. Data across all the trials are concatenated to obtain the correlation coefficients. This illustrates the utilization of a measure of similarity between channels of interest. Note that other similarity measures, or criteria different from the correlation coefficients, could be defined within this function or scheme [23,24].

The inputs to this function are as follows:

1. *eog_data*: Preprocessed data structure that contains only the channels corresponding to EOG data measures.
2. *data*: Preprocessed data structure with the EEG channels selected.

The output of this function is as follows:

1. *corrcoef_matrix_output*: Matrix with one row per EOG channel and one column per data channel with the corresponding correlation coefficients.

*4.3. Selection of Components to Remove: fta_select_comp_from_coeff_correlation*

This function selects outliers, within the set of correlation coefficients between the ICA components of EEG data and EOG data, to identify the channels to be rejected for EEG signal reconstruction. In addition, the coefficients are sorted in descending order of amplitude as a guide to the utilization of the other criteria.

The inputs to the function are as follows:

1. *corrcoef_ica_sum*: Contains the sum of the coefficients of the correlation matrix.

The output of the function is as follows:

1. *reject_comp*: Contains the selected components to remove.

Now, signal reconstruction can proceed with the selected data. Other criteria could be easily employed for the channel identification task.

*4.4. Reconstruction of EEG Channels from Independent Components: fta_channels_from_ica_reject_components*

This function reconstructs EEG channels from the ICA component structure, rejecting the selected ICs given by *fta_select_comp_from_coeff_correlation*. Original data (data before ICA decomposition) is required to match the data structure and vector lengths.

This function makes use of the following:

- *fta_select_comp_from_coeff_correlation*: Ad hoc function that selects the ICs to be rejected for EEG signal reconstruction (see Section 4.3).

The inputs to the function are as follows:

1. *ICA_comp*: Data structure as given by *ICA_comp = ft_componentanalysis (cfg_ica, data)*
2. *data_orig*: Original data before applying ICA.

3. *corrcoef_ica_original_sum*: Sum of correlation coefficients between the EOG channels and the rest of the EEG data channels.

The output of the function is as follows:

1. *data_reconstructed*: EEG data reconstructed from the accepted ICs.

*4.5. Creation of Artifact Membership Function: fta_create_msf_fieldtrip*

This function creates the artifact membership function (MSF) in the form of Fieldtrip structured data. The MSF is a logical sequence with the same length as the recorded data, in which artifacts are marked with +1 and other samples with 0.

The proposed function for illustration of the construction of the MSF uses the *ft_artifact_zvalue* function, which is based on the use of a Z-transform to automatically detect samples with different types of artifacts [17]. Nevertheless, other functions or algorithms could be used, such as wavelet-based schemes [12], as long as the final representation of the artifact detection follows the conventions in the provided examples. In the proposed example, to detect ocular artifacts, the following input parameters are used:

1. Channels: EOG (HEOG and VEOG).
2. Cut-off: Threshold of the Z-transformed values, above which the samples are considered as artifacts. A value of 0.5 is used.
3. Band-Pass Filter: Order 3 Butterworth filter between 2 and 15 Hz.
4. Artifact padding: Used to extend the temporal extent of artifacts on both sides. A value of 0.1 is used.
5. Trial padding: Used to include data segments before or after the trial. A value of 0 is used.

After obtaining the vector where samples with artifacts are marked, an MSF sequence with a Fieldtrip structure is constructed so that it fits the structure and the trial specifications of the EEG data.

This function makes use of the following:

- *ft_artifact_zvalue*: Scans data segments of interest for artifacts by means of thresholding the Z-transformed value using the pre-processing options to find ocular artifacts.

The inputs to the *fta_create_msf_fieldtrip* function are as follows:

1. *data_eeg*: EEG data to remove artifact influence.

The output of this function is as follows:

1. *MSF*: Logical sequence with Fieldtrip structure where artifacts are defined as 1 and other samples as 0.

*4.6. Partial Removal of Independent Components: fta_ica_partially_removed_components_method*

This function removes independent components from EEG signals with a high correlation with EOG signals, but only where artifacts were detected. For this, the correlation coefficients between ICA components and EOG data are used, and the impact of the components selected is removed at the samples marked as artifacts by the MSF.

The syntax to use with this function is as follows:

*cleaned_data = fta_ica_partially_removed_components_method(data_eeg, MSF, fig)*. This function partially removes ICs of EEG data using the MSF and correlation information, and draws (if *fig* = 1) a figure with the correlation between EEG and EOG signals before and after the procedure.

The steps for this method are as follows:

1. Obtain the VEOG and HEOG channels.
2. Obtain ICs of all the EEG channels.
3. Calculate the correlation coefficients between the ICs and the EOG signals, as in the previous method.
4. Sum all the correction coefficients between the EOG channels and each of the ICs.

5. Partial elimination of ICs related to ocular artifacts using the function *fta_channels_from_ic_partial_reject_components*.

    This function makes use of the following:

- *ft_selectdata* to select EOG data.
- *ft_componentanalysis* to obtain the ICs of all the EEG channels.
- *fta_corrcoeff*, which calculates correlation coefficients.
- *fta_channels_from_ica_partial_reject_components*, to partially reject components for reconstruction, where MSF = 1.

    The inputs to this function are as follows:

1. *data_eeg*: EEG data to process.
2. *MSF*: Logical sequence with Fieldtrip structure compatible with data_eeg, where artifacts are marked as 1 and other samples as 0, as created by *fta_create_msf_fieldtrip* (see Section 4.5).
3. *fig*: Logical value: 1 or 0 to plot, or not, the sum of the correlation coefficients before and after applying the method.

    The output of this function is as follows:

1. *cleaned_data*: Fieldtrip-structured data with reduced influence of ocular artifacts.

    In the examples section, a brief illustrative script of Matlab code will be shown.

*4.7. Reconstruction of EEG Channels from Independent Components Partially Rejected: fta_channels_from_ica_partial_reject_components*

This function reconstructs the EEG channels from selected ICs. The selected components, obtained depending on the sum of the correlation coefficients, are partially removed according to the MSF. Original EEG data (data before ICA decomposition) are required to match the data structure and vector lengths.

This function makes use of the following:

- *fta_select_comp_from_coeff_correlation*: Ad hoc function that selects the ICs to be rejected for EEG signal reconstruction (see Section 4.3).

    The inputs to this function are as follows:

1. *ICA_comp*: Data structure of ICs, as given by *ft_componentanalysis*.
2. *data_orig*: Original data before applying ICA.
3. *corrcoef_ica_original_sum*: Sum of the correlation coefficients between EOG channels and the rest of the channels (EEG data).
4. *MSF*: Logical sequence with Fieldtrip structure compatible with data_eeg, where artifacts are marked as 1 and other samples as 0, as created by *fta_create_msf_fieldtrip* (see Section 4.5).

    The output of this function is as follows:

1. *data_reconstructed*: EEG data reconstructed from ICs, where some of them were removed, or attenuated, at the time instants indicated by the MSF.

*4.8. Calculation of Artifact-Free Unmixing Matrix: fta_artifacts_free_ica_partially_removed_components _method*

This function calculates an artifact-free version of the ICA unmixing matrix to obtain ICs without artifact influence, making use of an MSF. The correlations between EOG data and independent components are used to remove these components at the samples marked as artifacts by the MSF.

This is the syntax to use this function:

*cleaned_data = fta_ artefacts_ free_ ica_ partially _removed _components _method (data_eeg, MSF, fig)*; this function also plots an illustration of the correlation between EEG and EOG signals before and after the process, when *fig* = 1.

The steps of this method are as follows:

1. Obtain VEOG and HEOG channels.

2. Obtain the unmixing matrix from ICA of the EEG data free of artifacts.
3. Recalculate the ICs of the complete EEG data using the artifact-free unmixing matrix calculated previously.
4. Calculate the correlation coefficients between the ICs and EOG signals, as in the previous method.
5. Perform the partial elimination of the ICs related to ocular artifacts using the correlation coefficients, as in the previous method.

This function makes use of the following:

- *ft_selectdata(cfg, data_eeg)* to select EOG data.
- *fta_ica_from_artifact_free_data* to obtain artifact-free ICs and the unmixing matrix.
- *ft_componentanalysis* to apply ICA using the artifact-free unmixing matrix.
- *fta_corrcoeff* to calculate the correlation coefficients.
- *fta_select_comp_from_coeff_correlation* to select the ICs with a high correlation with EOG signals.
- *fta_channels_from_ic_partial_reject_components* to partially reject ICs, where MSF = 1.

The inputs to this function are as follows:

1. *data_eeg*: EEG data with artifacts.
2. *MSF*: Logical sequence with Fieldtrip structure compatible with *data_eeg*, where artifacts are marked as 1 and other samples as 0.
3. *fig*: Logical value that takes the value of 1 to plot the sum of the correlation coefficients, before and after applying the method.

The output of this function is as follows:

1. *clean_data*: Fieldtrip-structured data with the influence of ocular artifacts diminished by partially removing artifacts using an artifact-free ICA unmixing matrix and MSF.

In the examples section, a small illustrative excerpt of Matlab code will be shown.

*4.9. Independent Component Analysis with Artifact-Free Data: fta_ica_from_artifact_free_data*

This function performs ICA, making use of *ft_componentanalysis* with all the data samples not marked as artifacts by the MSF.

This function makes use of the following:

- *ft_componentanalysis*: To perform ICA on EEG data.

The inputs to this function are as follows:

1. *data*: EEG data channels, as obtained after initial pre-processing.
2. *MSF*: Logica sequence with Fieldtrip structure compatible with *data_eeg*, where artifacts are marked as 1 and other samples as 0.

The output of this function is as follows:

1. *comp_ica_from_artifact_free_ica*: Reconstructed ICA data with partial removal of the ICs related to artifacts.

## 5. Examples

In this section, illustrative examples of the results obtained by using the methods and code proposed are shown to facilitate the reader to follow and replicate the processes.

The figures obtained, as well as the data used in the examples shown, correspond to the processing of EEG data of a certain subject (denoted subject 001), recorded according to the procedure, and the experiment described in Section 2.

**Method 1: ICs removal.** ICA components with a high absolute value of their correlation with EOG signals are simply eliminated; as a consequence, a significant portion of information can be lost.

In Figure 1, three graphs can be observed: Figure 1a represents EOG signals, Figure 1b shows the Fp1 channel (which usually appears strongly affected by ocular artifacts) and Fz of the original EEG signal, and, in Figure 1c, the same channels for the decontaminated EEG signals from the proposed method are shown.

In Figure 1c, the observable large peaks linked to ocular artifacts have been removed. In addition, it can be observed that in some areas not influenced by artifacts there is a decrease in amplitude with respect to the original shape of the signal.

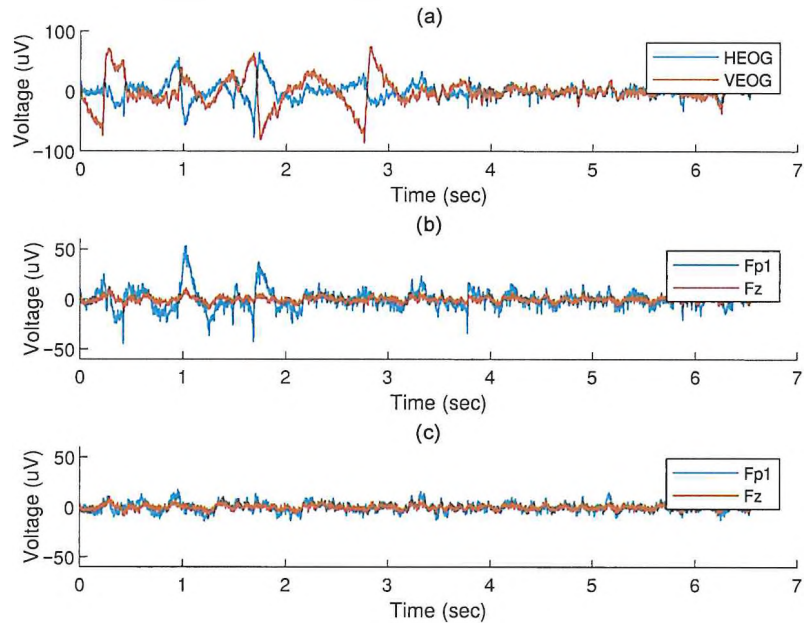

**Figure 1.** Method 1: Removal of ICs. (**a**) HEOG and VEOG signals; (**b**) original EEG data (for channels Fp1, near the eyes, and Fz); and (**c**) decontaminated data obtained after artifact removal with the proposed method.

An illustrative example of the utilization of this method is shown in Box 1.

**Box 1.** Matlab example code for the utilization of the proposed function for the complete elimination of ICs.

```
%%EEG file
filename = 'SUJ_001_EXP_001_1.eeg';
%%Preprocessing EEG files
cfg = [];
cfg.dataset = filename;
%Remove the audio
cfg.channel = {'all', '-Audio'};
%Reference channel
cfg.refchannel = 'FCz';
%Correction by the reference point
cfg.demean = 'yes';
cfg.detrend = 'yes';
cfg.lpfilter = 'yes';
cfg.lpfreq = 250;
cfg.lpfilttype = 'fir';
cfg.lpfiltdir = 'twopass';
cfg.hpfilter = 'yes';
cfg.hpfreq = 1;
cfg.hpfilttype = 'fir';
cfg.hpfiltdir = 'twopass';
data_eeg = ft_preprocessing(cfg);
%%FIRST METHOD
fig = 1;
clean_data = fta_ica_removed_components_method(data_eeg, fig);
%Visualize the clean EEG data
ft_databrowser([],clean_data);
```

The reduction of the correlation coefficients between the independent components and EOG data after using the method is illustrated in Figure 2; specifically, Figure 2b shows a significant reduction of the correlation coefficients with respect to Figure 2a, after the application of the procedure.

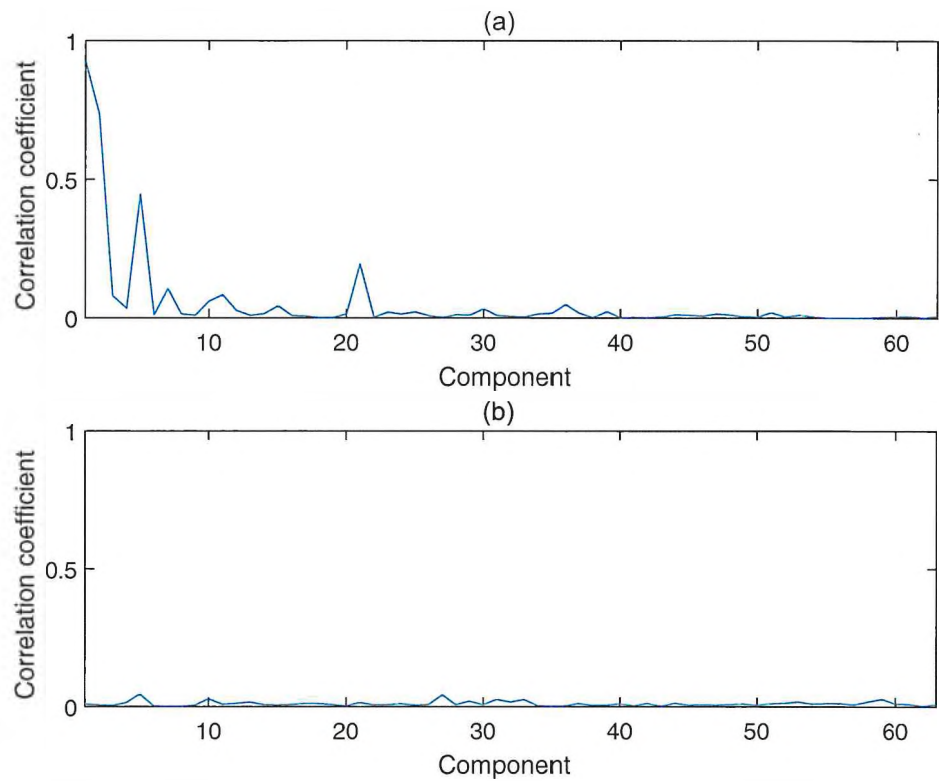

**Figure 2.** Sum of the correlation coefficients between ICs and EOG data (**a**) before and (**b**) after the rejection of the highest correlation components in the example of Method 1.

The identification of the ICs to remove is implemented as a well-separated procedure, which should allow the utilization of other methods based on schemes such as the ones described in [23,24], or others.

**Method 2: Partial removal of ICs.** This method requires the identification of specific samples in which artifacts are present. This artifact detection process can be performed by hand or by automatic means to build the MSF (see Section 3). This method might somehow resemble PCA decomposition in sliding windows, as used in the ASR schemes for artifact removal [8]; however, the manner in which the windowing procedure is considered, and the way in which data are employed, are completely different.

Figure 3 shows signal excerpts with samples where artifacts have been automatically detected, and marked, building the MSF shown in red. Then, the decontamination process is carried out similarly to Method 1, but only for the samples in which artifacts were found. The signal obtained after the process is expected to be without artifacts; additionally, less information should be lost in the rest of the samples, as compared to the previous method.

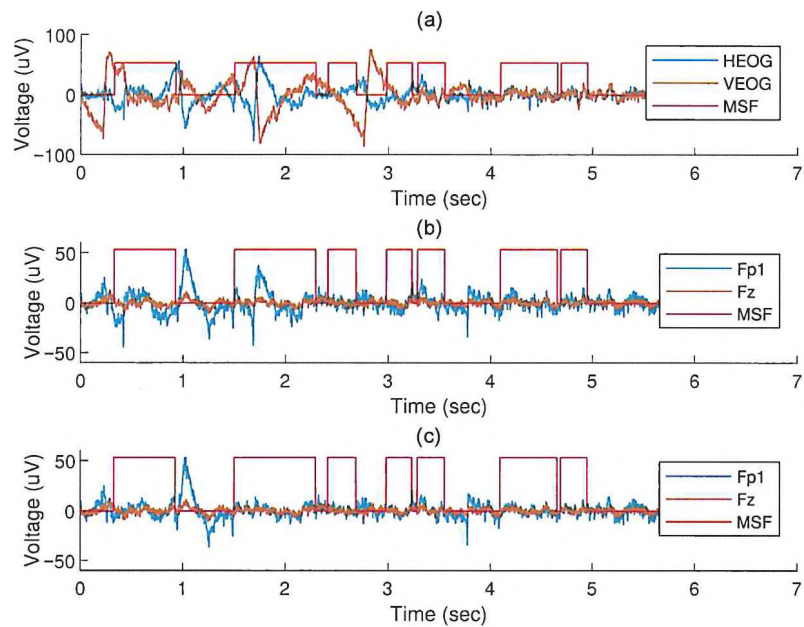

**Figure 3.** Method 2: Partial removal of ICs. (**a**) HEOG and VEOG signals, together with artifact detection recorded in the MSF; (**b**) original EEG data (for the Fp1 channel, near the eyes, and Fz); and (**c**) data obtained after artifact removal with the proposed method (also for the Fp1 and Fz channels).

An example of the usage procedure of this method is shown in Box 2.

**Box 2.** Matlab example code for the application of the proposed function for partial elimination of ICs using the MSF.

```
%%EEG file
filename = 'SUJ_001_EXP_001_1.eeg';
%%Preprocessing EEG files
cfg = [];
cfg.dataset = filename;
%Remove the audio
cfg.channel = {'all', '-Audio'};
%Reference channel
cfg.refchannel = 'FCz';
%Corretion by the reference point
cfg.demean = 'yes';
cfg.detrend = 'yes';
cfg.lpfilter = 'yes';
cfg.lpfreq = 250;
cfg.lpfilttype = 'fir';
cfg.lpfiltdir = 'twopass';
cfg.hpfilter = 'yes';
cfg.hpfreq = 1;
cfg.hpfilttype = 'fir';
cfg.hpfiltdir = 'twopass';
data_eeg = ft_preprocessing(cfg);
%MSF creation
MSF_ft = fta_create_msf_fieldtrip(data_eeg);
%%SECOND METHOD
fig = 0;
cleaned_data = fta_ica_partially_removed_components_method(data_eeg, MSF_ft,fig);
%Visualize the cleaned EEG data
ft_databrowser([],cleaned_data);
```

**Method 3: Partial removal of ICs using artifact-free ICA unmixing matrix.** In this case, a different unmixing matrix is found by using solely uncontaminated samples; to this end, an MSF is again required. An example of the results of the utilization of this method is shown in Figure 4. Note that this figure shows the same signal and MSF excerpts as in the previous cases, for comparison purposes; differences are clearly observable in the results with respect to those methods.

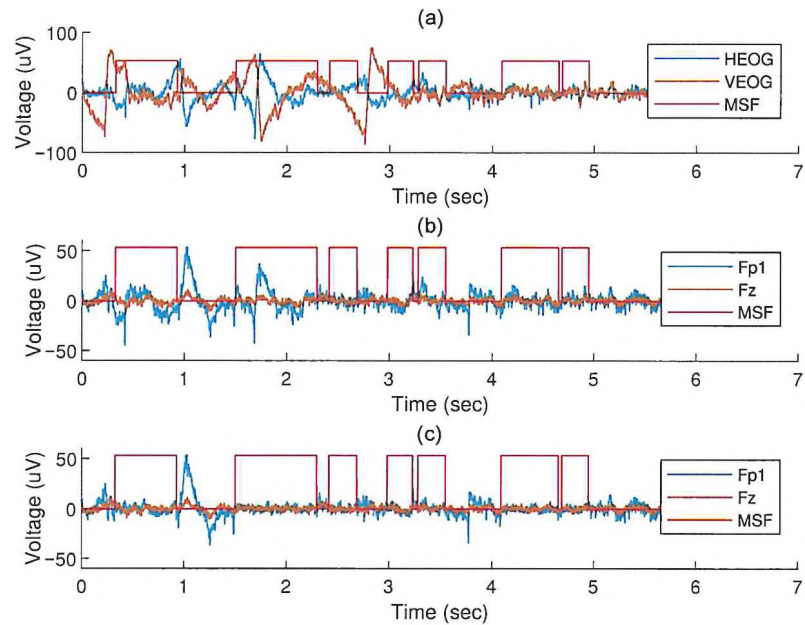

**Figure 4.** Method 3: Partial removal of ICs using artifact-free unmixing matrix. (**a**) HEOG and VEOG signals, together with artifact detection results expressed in the MSF; (**b**) original EEG data (for channels Fp1, near the eyes, and Fz); and (**c**) data obtained after artifact removal with the proposed method.

An example of Matlab code to demonstrate the utilization of this method is provided in Box 3.

**Box 3.** Matlab example code for the application of the proposed function for the partial elimination of ICs using artifact-free unmixing matrix and MSF.

```
%%EEG file
filename = 'SUJ_001_EXP_001_1.eeg';
%%Preprocessing EEG files
cfg = [];
cfg.dataset = filename;
%Remove the audio
cfg.channel = {'all', '-Audio'};
%Reference channel
cfg.refchannel = 'FCz';
%Corretion by the reference point
cfg.demean = 'yes';
cfg.detrend = 'yes';
cfg.lpfilter = 'yes';
cfg.lpfreq = 250;
cfg.lpfilttype = 'fir';
cfg.lpfiltdir = 'twopass';
cfg.hpfilter = 'yes';
cfg.hpfreq = 1;
cfg.hpfilttype = 'fir';
cfg.hpfiltdir = 'twopass';
data_eeg = ft_preprocessing(cfg);
%MSF creation
MSF_ft = fta_create_msf_fieldtrip(data_eeg);
%%THIRD METHOD
fig = 0;
cleaned_data = fta_artefacts_free_ica_partially_removed
_components_method(data_eeg, MSF_ft,fig);
%Visualize the cleaned EEG data
ft_databrowser([],cleaned_data);
```

With these examples, this manuscript, and the associated data and code, responds to some of the issues highlighted in [13] by providing all the components necessary for the exact replication of the procedures, along with the experiments presented.

## 6. Conclusions

A Matlab–Fieldtrip implementation of tools aimed to reduce the influence of artifacts has been described and made available. Experiments, data, and sample code specifically focused on eye movement artifacts are shown. Code scripts, functions, and sample data are made available to the researchers to facilitate and foster the utilization and further development of the techniques shown. The methods are implemented using a FieldTrip toolbox, though the specific choices and processing schemes can be implemented using other software tools or programming languages for their integration into other systems.

A functionality from the Fieldtrip toolbox is used to identify artifacts; however, other methods could also be used. These could be based on diverse measures of similarity between the signals' time-varying energy, or signal derivatives, or other features. The definition of the artifact membership function is kept very simple to make its creation straightforward by using other artifact detection schemes. In addition, the MSF is employed in this work as a logical sequence; however, it can be easily turned into a weighting function to control the strength of the corresponding artifact removal processes shown.

Regarding the selection of ICs to perform the processes for the removal of the influence of artifacts, other measures different from the correlation coefficients could be considered. The code provided can be easily adapted to other measures for this selection task.

Researchers are encouraged to use, modify, and improve the techniques described by employing the sample code provided at https://hdl.handle.net/10630/26154, accessed on 16 January 2024.

**Author Contributions:** Conceptualization, M.M.-M., A.M.B., L.J.T. and I.B.; methodology, M.M.-M., A.M.B., L.J.T. and I.B.; software, M.M.-M., A.M.B. and L.J.T.; validation, A.M.B., L.J.T. and I.B.; formal analysis, M.M.-M., A.M.B., L.J.T. and I.B.; investigation, M.M.-M., A.M.B., L.J.T. and I.B.; resources, I.B.; data curation, M.M.-M., A.M.B. and L.J.T.; writing—original draft preparation, M.M.-M., A.M.B., L.J.T. and I.B.; writing—review and editing, M.M.-M., A.M.B., L.J.T. and I.B.; visualization, M.M.-M.; supervision, L.J.T. and I.B.; project administration, I.B.; funding acquisition, I.B. All authors have read and agreed to the published version of the manuscript.

**Funding:** This publication is part of the project PID2021-123207NB-I00 funded by MCIN/AEI/10.13039/501100011033/FEDER,UE.

**Institutional Review Board Statement:** Data acquisition methodology was conducted according to the guidelines of the Declaration of Helsinki and approved by the Comité de Ética de la Investigación Provincial de Málaga on 19 December 2019, approval number: 2176-N-19.

**Informed Consent Statement:** Informed consent was obtained from all subjects involved in the study.

**Data Availability Statement:** Data and sample code are provided at https://hdl.handle.net/10630/26154 (accessed on 16 January 2024), researchers are encouraged to use, modify, and improve the techniques described.

**Conflicts of Interest:** The authors declare no conflicts of interest.